\advance\topmargin by 2.0in
\documentclass[aps,prl,twocolumn]{revtex4}
\usepackage{graphicx}
\usepackage{epsfig}

\begin{document}

\title{Magnetization plateaus for spin-one bosons in optical lattices:
Stern-Gerlach
experiments with strongly correlated atoms}

\author{Adilet Imambekov, Mikhail Lukin, and Eugene Demler}
\affiliation{Department of Physics, Harvard University, Cambridge
MA 02138}

\date{\today}

\begin{abstract}

We consider insulating states of spin-one bosons in optical lattices in the
presence of a weak magnetic field. For the states with more than one atom per
lattice site we find a series of quantum phase transitions between states with
fixed magnetization and a canted nematic phase.  In the presence of a global
confining potential, this unusual phase diagram leads to several novel
phenomena, including formation of magnetization plateaus. We discuss how these
effects can be observed using spatially resolved density measurements.

\end{abstract}

\maketitle Far-off-resonance optical traps can confine neutral atoms regardless
of their hyperfine state \cite{MITOpticalTrapping} and open exciting
possibilities for exploring multicomponent systems of degenerate atoms
\cite{Ho98,OhmiMachida,duan2003,HoYip,japan,KoashiUeda2000,Law98,pra,snoekzhou,svidzinskychui,zhou1,zhou2,svistunov,japan2,DemlerZhou}.
Experimental studies of the hyperfine spin-one manifold of $^{23}$Na atoms
confined in a single optical trap already revealed the ground state spin
structure in external magnetic field \cite{MitNature}, the formation and
persistence of the metastable spin domain configurations \cite{mitdomains}, and
quantum tunnelling across spin domain boundaries \cite{mitbectunnelling}. These
phenomena have been understood using mean-field theory
\cite{Ho98,OhmiMachida,LESHouchesreview} since in a large trap atoms can be
considered as weakly interacting. At the same time, theoretically predicted
many-body features of spin-one condensates, such as a spin singlet nature of the
ground state for even number of atoms \cite{Law98}, could not be observed in
such experiments, since the energy difference between various spin states is
inversely proportional to the volume of the system and is extremely small for
realistic traps \cite{HoYip,KoashiUeda2000}.

Several approaches have recently been suggested to creating strongly
correlated quantum states of spinor atoms  in optical lattices
produced by standing wave laser fields
\cite{DemlerZhou,svistunov,duan2003,pra,snoekzhou}. In the
presence of a deep optical lattice wavefunctions of atoms are
localized near the lattice potential minima, which leads to a
strong enhancement of interactions between atoms
\cite{Jaksch98,Bloch,Orzel} and can result in dramatic changes in
the properties of ultracold gases.  It is also important that in
optical lattices one can have a small number of atoms per lattice
site (in experiments of Ref. \cite{Bloch} this number was around
1-3). In an insulating state  hopping between different sites is
negligible and each well behaves essentially as a small condensate.
In this regime, the behavior of  spin-one bosons in each well will
be dominated by the strong spin-spin interactions such as those
predicted by \cite{Law98} . Two component atom mixtures in optical
lattices have already been realized in experiments of Ref.
\cite{bloch2}.

Non-trivial nature of the ground state of localized spin-one bosons in optical
lattices has been discussed previously in \cite{DemlerZhou,pra,snoekzhou}.
Virtual hoppings of the atoms give rise to spin exchange interactions between
neighboring sites and lead to several distinct insulating phases, which differ
in their spin correlations.  In two and three dimensional lattices states with
odd numbers of particles per site are always nematic, and for states with even
numbers of particles per site, there is always a spin singlet phase, and there
may also be a first order transition into the nematic phase. In this paper we
extend earlier analysis and show that in a magnetic field insulating states
with more than one atom per site undergo a series of phase transitions between
spin gapped phases with quantized magnetization and phases with so called
canted nematic order, in which magnetization can vary continuously. The
critical properties of these phase transitions have been recently investigated
in \cite{zhouaffleck}. Magnetic phase diagram is similar to the phase diagram
of the spinless boson Hubbard model: states with quantized magnetization are
analogous to the Mott insulating phases, while the canted nematic phase is
analogous to the superfluid phase. We also propose two kinds of experiments
that can verify the presence of such magnetization plateaus and demonstrate the
many-body nature of insulating states of spin-one bosons in optical lattices.

An effective Hamiltonian for spin-one bosons in an optical lattice in
the presence of a magnetic field is given by \cite{pra}
\begin{eqnarray}
{\cal H}=-t\sum_{\langle i j\rangle,\sigma}(a^{\dagger}_{i
\sigma}a_{j \sigma}+a^{\dagger}_{j \sigma}a_{i \sigma}) +
\frac{U_0}{2}\sum_{i}\hat n_i(\hat n_i-1) \nonumber \\
 + \frac{U_2}{2}
\sum_{i}(\vec S_i^2-2 \hat n_i ) - \mu \sum_{i}\hat n_i-H\sum_i
S_{z i}, \label{OriginalHamiltonian}
\end{eqnarray}
Here $a^{\dagger}_{i\sigma}$ are creation operators for particles
in the lowest Bloch band localized on site $i$ and having spin
components $\sigma=0,\pm 1$; $n_i= \sum_\sigma a^{\dagger}_{i
\sigma} a_{i \sigma}$ and $\vec S_i = \sum_{\sigma \sigma'}
a^{\dagger}_{i \sigma} \vec{T}_{\sigma \sigma'} a_{i \sigma'}$ are
the number and spin operators for site $i$ ($\vec{T}_{\sigma
\sigma'}$ are the usual spin matrices for spin 1 particles). For
each well the collective spin of the atoms satisfies constraints $
S_i \leq N_i $ and $S_i + N_i$ is even. Parameters $t, U_0$ and
$U_2$ for a realistic case of a three dimensional cubic lattice
 have been obtained in \cite{pra}. The ratio of the interaction terms in
(\ref{OriginalHamiltonian}), $U_2/U_0$, is fixed by the ratio of
the scattering lengths and is independent of the nature of the
lattice. Scattering lengths of $^{23}Na$ obtained in Ref.
\cite{Nascatteringlengts} give $U_2/U_0=0.04.$
In this paper we neglect effects of the quadratic Zeeman shift
since magnetization plateaus that we are interested in appear for
magnetic fields of the order of mGauss (assuming a typical
$U_2\approx 0.1 {\rm kHz}$).

\begin{figure}
\psfig{file=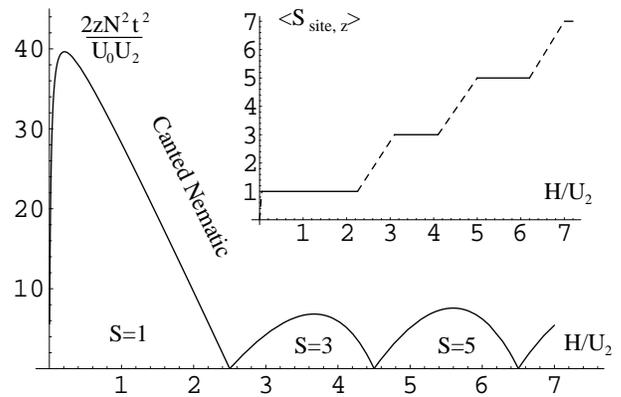} \caption{\label{oddS} Magnetic insulating phase of spin-one
bosons in optical lattice with an { \it odd} number of atoms per site $N$.
Magnetization is fixed inside the lobes but varies continuously inside the
Canted Nematic phase. The latter has an expectation value of the nematic order
parameter $Q_{ab}$ in the plane perpendicular to the direction of magnetic
field. For sufficiently large magnetic field the system becomes fully polarized
with $S=N$(not shown here). The insert shows the ground state magnetization
(per site) as a function of magnetic field for $2zN^2t^2/U_0=5 U_2$. Note, that
near $H=0$ magnetization grows linearly with magnetic field and quickly reaches
the $S_{\rm site,z}=1$ plateau.}
\end{figure}

\begin{figure}
\psfig{file=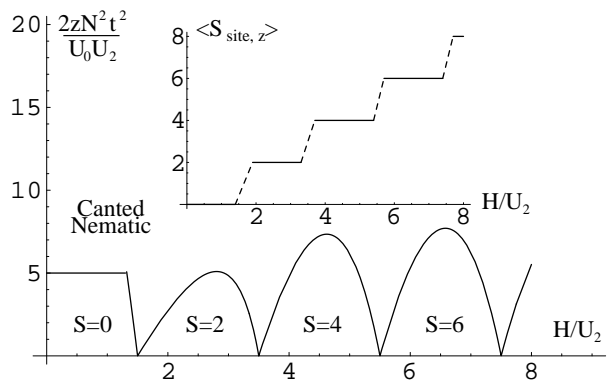} \caption{\label{evenS} Magnetic insulating phase in
optical lattice with an { \it even} number of atoms per site. The insert is for
$2zN^2t^2/U_0=2 U_2.$}
\end{figure}

When the spin dependent interaction ($U_2$) is much smaller than
the Hubbard repulsion ($U_0$), the superfluid - insulator
transition\cite{Fishers89,Jaksch98} is determined mostly by $U_0$.
The spin gap $U_2$ term, however, is important inside the
insulating regime, in which nontrivial spin phases appear as a
result of a competition between a spin gap, a magnetic field, and
spin exchange interactions, induced by fluctuations in the
particle number.  The magnetic phase diagram can be
most easily understood by considering the limit of large number of
atoms per site, $N\gg 1$. In this case the effective spin
Hamiltonian for the insulating state can be written as a model of
quantum rotors, interacting via rotationally invariant quadrupolar
interaction \cite{pra,snoekzhou}:
\begin{eqnarray}
{\cal H} &=& \sum_i\frac{U_2}{2}  \vec{S}_i^2 - H S_{zi} -\frac{2N^2t^2}{U_0}
\sum_{ij} n_{ia} n_{ib}n_{ja}n_{jb}.
 \label{NbigEffectiveHamiltonian}
\end{eqnarray}
Angular momentum $\vec{S}_i= -i\, {\bf n}_i \times
\frac{\partial}{\partial {\bf n}_i}$ describes the collective spin
on site $i$ and inherits constraints of the microscopic
Hamiltonian (\ref{OriginalHamiltonian}).
In the mean field approximation \cite{QPT,SS} we replace
(\ref{NbigEffectiveHamiltonian}) by a sum of single site
Hamiltonians
\begin{eqnarray}
H_{MF,i}= \frac{U_2}{2}  \vec{S}_i^2
 - H S_{zi}- \frac{zN^2t^2}{U_0} \,\,
(Q_{ab}+\frac{1}{3}\, \delta_{ab}) \,\,
n_{ia}n_{ib}
\label{hmf}
\end{eqnarray}
with $z$ being the number of nearest neighbors, and impose a self-consistency
condition on the nematic order parameter $Q_{ab} = \langle n_{ia} n_{ib}\rangle
-\frac{1}{3}\, \delta_{ab}$. Magnetic phase diagrams obtained by solving
(\ref{hmf}) self-consistently for the cases of odd and even filling factors are
shown in Figs. \ref{oddS} and \ref{evenS} (for details see \cite{magn_theory}).
When $t=0$ the Hamiltonian (\ref{hmf}) does not mix different spin eigenstates.
In a magnetic field the system has a series of level crossings between states
with different values of the spin. Each of these states has a gap in the
excitation spectrum and remains stable after turning on a finite value of $t$.
This results in lobes of fixed magnetization shown in Figs. \ref{oddS} and
\ref{evenS}.  Only at points where spin eigenstates come into degeneracy
exchange interactions give rise to mixing of different spin eigenstates,
leading to a Canted Nematic phase. The latter has an expectation value of the
nematic order parameter $Q_{ab}$ in a plane perpendicular to the magnetic field
\cite{zhouaffleck}. Hence, it has spontaneous breaking of the symmetry of spin
rotations around the direction of magnetic field. For sufficiently large
magnetic field the system becomes fully polarized with $S_i=N$. We note that
for $N=1$ and $N=2$ one can derive effective spin interactions that do not rely
on the large $N$ approximation of equation (\ref{NbigEffectiveHamiltonian})
\cite{pra}. These models give magnetic phase diagrams that are qualitatively
similar to Figs. \ref{oddS} and \ref{evenS} \cite{magn_theory}. We also point
out that inside the lobes of fixed magnetization and in the mean-field
approximation the many-body wavefunctions factorize $ |\Psi \rangle = \prod_i
|S_i=S,S_{iz}=S \rangle $.


A typical experimental system has a parabolic confining potential in addition
to the periodic optical lattice. This means that the chemical potential is not
uniform and for small $t$ the system breaks into insulating domains with
different integer filling factors, separated by  regions of the superfluid
phase (see Refs. \cite{Jaksch98,numerics} for analysis of the spinless case).
Another important experimental constraint is that the total magnetization is
fixed by the initial state of the system. In the discussion above we showed,
that when we change magnetic field, it is energetically favorable to adjust the
magnetization in order to utilize some of the Zeeman energy (see Figs.
\ref{oddS} and \ref{evenS}). From equation (\ref{OriginalHamiltonian}) we
observe, however, that the spin component parallel to the applied field is
conserved.  For example, if the magnetic field is along the $z$ axis, $S_{\rm
tot,z}=\sum_i S_{iz}$ is a good quantum number of the system (spin
non-conserving interactions, such as the dipolar relaxation, are typically
small) and should not change even when $H_z$ is changing.  For a single large
trap this feature allowed to study magnetic properties of spinor
condensate\cite{MitNature} at magnetic fields for which the true ground state
should be fully polarized.

\begin{figure}
\psfig{file=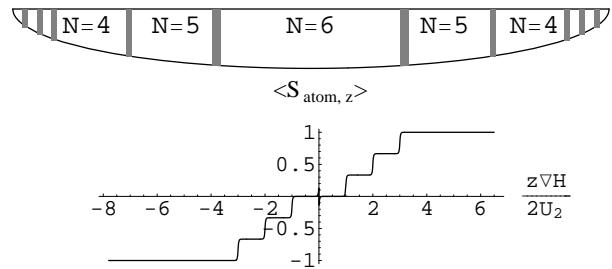} \caption{\label{cond_plateaus}
A system of spin-one bosons in an
optical lattice confined by a parabolic potential.
In the insulating regime the cloud breaks into
insulating domains with different integer filling factors.
When magnetic field gradient is applied parallel to the long axis
of the trap, magnetization plateaus  develop inside individual
insulating domains.}
\end{figure}

A way around spin conservation has been demonstrated in
\cite{MitNature} and relies on applying spatially varying magnetic
fields and performing Stern-Gerlach imaging. Here we extend these
ideas and suggest an approach to experimental observation of the
magnetization plateaus discussed above. The idea of our first
experiment is shown in Fig. \ref{cond_plateaus}.
We consider a strongly anisotropic trap in which a magnetic field
gradient is applied parallel to the long axis of the system.
When there is a magnetic field gradient parallel to
the long axis, there should also be gradients in the transverse directions
(both $\vec{\nabla}\vec{H}$ and $\vec{\nabla}\times \vec{H}$ should be zero).
We assume that the size of the condensate in the transverse directions is small
enough that we can neglect  the effects of the magnetic field in transverse
directions. However, the size of the condensate should be larger than the
optical lattice period in any direction, so that we can consider it as a three
dimensional system.  As we discussed before, the uniform part of the magnetic
field has no effect on the state of the system, so in our discussion we set it
to zero. To be concrete, we assume that the largest insulating domain at the
center of the trap has a filling factor $N=6$ and that our system has been
prepared to have $S_{\rm tot,z}=0$. In a nonuniform magnetic field different
parts of the $N=6$ domain minimize their energy for different values of
magnetization. When the field gradient is sufficiently large and tunnelling is
small enough, the locally favored magnetization changes in a step like fashion
from $S_{\rm site,z}=-6$ (per site) on the left to $S_{\rm site,z}=6$ on the
right. Such state is also consistent with the spin conservation, since $S_{\rm
tot,z}$ remains zero. So, the configuration that minimizes the energy has
plateaus in the spatial profile of magnetization, with spin polarizations (per
atom) $S_{\rm atom,z}=-1,-2/3,-1/3,0,1/3,2/3,1,$  as shown in Fig.
\ref{cond_plateaus}. Each plateau has a length of the order of $2 U_2/|\nabla
H|$. The appearance of spin plateaus in a nonuniform magnetic field is
analogous to the domain structure of condensates in optical lattices in the
presence of a non-uniform global confining potential that was discussed before
\cite{Jaksch98,numerics}. In the latter case the total density is fixed by the
number of atoms in the trap, but insulating phases with different integer
filling factors exist due to the confining potential. To detect the
magnetization plateaus shown in Fig. \ref{cond_plateaus}, one needs to image
different parts of the trap separately, measuring spin polarization per atom as
a function of the position: $\langle S _{\rm atom,z}\rangle = (n_+ - n_-)/(n_+
+ n_0+ n_-)$, where $n_\pm, n_0$ are the local densities of atoms with
$\sigma=\pm1 $ and $0$ respectively.  This quantity can be most easily measured
in Stern-Gerlach time-of-flight experiments. If a small gradient of magnetic
field is applied during expansion, clouds with different spin components
spatially separate, and one can measure the number of atoms with different spin
components using light scattering.

To illustrate system parameters needed to realize this experiment we consider a
cigar shaped condensate of sizes $400 \times 10 \times 10$ $\mu$m.  For an
optical lattice created with $\lambda=985 $ nm lasers and $4\times 10^5$ atoms
in a condensate we get the maximum density of six atoms per well in the center
of the trap. To observe five plateaus in this setup, one would need magnetic
field gradients $\sim 100 {\rm mG/cm},$ and spatial resolution of $~30 {\rm \mu
m}$. These parameters have already been achieved in experiments of Ref.
\cite{MitNature}. Inside the insulating phase the characteristic timescale for
spin relaxation between different wells is set by the exchange interactions
$\tau_{ex}=\hbar U_0/(Nt)^2$. For $t=0.1 {\rm khz}$ and $U_0=2{\rm khz}$ we
find times of the order of hundreds of milliseconds. So, if a magnetic field
gradient is applied in the insulating regime, one needs to wait at least
that long for magnetic plateaus to develop.  Experimentally it may be more
efficient to apply magnetic field gradient when the system is in the superfluid
regime and then take the system to the insulating state by slowly reducing $t$.
\begin{figure}
\psfig{file=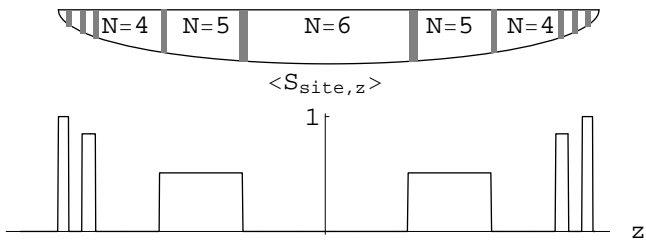} \caption{\label{modulation} A system
of spin-one bosons with a nonzero total magnetization in an optical
lattice (no external magnetic field).  Magnetization gets
distributed among regions with odd filling factors and is pushed out
of the regions with even fillings.}
\end{figure}

We now discuss our second "spin decoration" approach to experimental detection
of spin gap effects in the insulating regime of spin-one bosons. The idea of this
method is that when a system is prepared with a non-zero magnetization in the absence of magnetic field gradients,
magnetization gets distributed non-uniformly among insulating domains with
different filling factors. Spin polarization appears predominantly  in domains
with odd numbers of atoms per site and is pushed out of the regions with even
fillings. To justify this conjecture we propose the following argument.  
In the regime, when tunneling is small($zN^2t^2/U_0\ll U_2$), there is a crucial difference in spin susceptibility
between odd and even phases. For even domains, one needs to pay an energy $3 U_2$ to break a singlet state and have
$S_{ \rm site,z}\sim 2.$ For odd sites the lowest energy state already has $S=1,$ and energy cost to polarize existing spins 
is of the order of $zN^2t^2/U_0.$ Therefore, externally imposed nonzero magnetization will be redistributed in odd 
insulating domains. For small magnetization per site, energy goes as 
\begin{eqnarray}
E_{\rm odd}(N,S_{\rm site,z})=
\frac{1}{2\chi_{\rm odd}(N)} S^2_{\rm site,z}.
\label{EofMagnetizationOdd}
\end{eqnarray}
 If $\chi_{\rm odd}(N)$ was the same for all $N$, then magnetization would be
distributed uniformly among all odd domains. In reality, $\chi_{\rm odd}(N)$
decreases with increasing $N$, so we expect larger magnetization for insulating
domains with smaller number of atoms. Quadratic dependence in
(\ref{EofMagnetizationOdd}) ensures, however, that all domains with odd filing
factors acquire finite magnetization. So, in experiments we expect to find a
picture of alternating even and odd domains, in which odd domains have finite
magnetization and even domains have none(see Fig. \ref{modulation}). This
picture is valid until all odd regions have magnetization $S_{\rm site,z}=1.$
For the experimental setup discussed earlier this corresponds to $\langle
S_{atom,z}\rangle \lesssim 0.1.$ Performing spatially resolved measurements of
spin polarization one should be able to observe such a modulated structure of
magnetization.

In summary, we discussed insulating states of spin-one bosons in
optical lattices in the presence of a magnetic field. For systems
with more than one particle per site we demonstrated the existence
of a series of phase transitions between phases with fixed
magnetization and the canted nematic phase in which magnetization
can vary continuously.  We considered experimental signatures of
this novel magnetic phase diagram, including formation of
magnetization plateaus in the presence of a magnetic field
gradient and a modulated spin density in nonuniform systems with
non-zero total magnetization.
We thank E. Altman, D. Petrov, A. Sorensen, D.W. Wang, and A.
Polkovnikov for useful discussions. This work was partially
supported by NSF (PHY-0134776,DMR-0132874), Sloan
and Packard Foundations.

\end{document}